\documentstyle[12pt]{article}
\def\eqref#1{(\ref{#1})}
\def\hx{\hat x}
\def\hy{\hat y}
\def\hrho{\hat \rho}
\def\ri{i}

\begin{document}

\title{\vspace*{-2cm}\hfill {\normalsize UT-KOMABA 01-06}\\
\vspace{2cm}
\bf Variational Approach to Dynamics of Quantum Fields
\footnote{Lectures delivered at the NATO Advanced Study Institute on
{\it QCD Perspectives on Hot and Dense Matter}, Carg\`{e}se, Aug. 6 -
18, 2001.}} 
\author{T. MATSUI \\ {\it Institute of Physics, University
of Tokyo, Komaba}\\ {\it 3-8-1 Komaba, Meguro-ku, Tokyo 153-8902,
Japan}} \date{October 30, 2001}

\maketitle

\def\mib#1{\mbox{\boldmath $#1$}}

\section{Introduction}
In the summer of 1995, Dominique Vautherin came to Kyoto and stayed
three months at the Yukawa Institute of Kyoto University where he
delivered a series of very informal, pedagogical lectures on the
application of variational methods to quantum field
theories. \cite{DV95} This initiated our long lasting enjoyable
collaboration on the subject and Yasuhiko Tsue joined the force later.
Throughout this collaboration Vautherin was always enthusiastic to
explore new physics problems and always came up with new innovative
ideas, much of which I suspect have origins in his expertise in
nuclear many-body theory.  We are very sorry that this fruitful
collaboration with Vautherin came to end so soon by his untimely death
and that we are no longer able to be inspired by his deep physics
insights, charmed by his elegance, and most of all cheered up by his
very warm presence.  I like to dedicate these lectures to the memory
of Dominique Vautherin from whom I learned most of the material
presented below.

First, I like to say a few words about physics motivations of our
works.  In recent years, several authors have constructed non-trivial
time-dependent solutions of classical field equations of effective
meson fields. \cite{AR91,BK92} Such solutions are relevant in
considering the fate of defects which might be produced in dynamical
order-disorder phase transitions in ultrarelativistic nucleus-nucleus
collisions\cite{RW93} or in the evolution of very early
universe.\cite{GP85} We like to study {\sl the effect of quantum
fluctuations} which has been ignored in these classical analyses.  We
may consider the classical solution as a collective mode of underlying
microscopic degrees of freedom.  It is well-known in nuclear many-body
theory that the internal microscopic motions of nucleons are
influenced by the motion of the nuclear mean field such as rotation or
vibration.  One may expect similar effects would arise in the field
theory.

Our method is based on the Schr\"{o}dinger wave functional
representation of the quantum field theories in which the
time-evolution of the wave function is explicitly considered.  Another
advantage of using the Schr\"{o}dinger picture is that one can
introduce the variational method to construct approximate but
non-perturbative solutions of the problem as in similar problems in
quantum mechanics.  The extension of the variational principle for
time-dependent wave function has also been developed. \cite{JK79,BV81}
The method can also be extended to deal with statistical ensembles
described by Gaussian form of density matrices. \cite{EJP88}

\section{A Simple Exercise in Quantum Mechanics}

We start with a very elementary example in quantum mechanics which all
students learn in an introductory course of quantum mechanics: the
harmonic oscillator problem.  Although our interests lies in the
time-dependent solution of quantum field theories, it would be
instructive to first study the time-dependent solution of this simple
exactly soluble problem in order to gain some physical insights into
our more difficult problems since quantum field theories are nothing
but an assemble of infinite number of coupled harmonic oscillators.


Let us consider a simple one-dimensional harmonic oscillator whose
Hamiltonian is given by
\begin{equation}\label{ho}
H_0 = \frac{1}{2} \left( p^2 + \omega ^2 x^2 \right) ,
\end{equation}
where we set the mass of the particle $m=1$ for simplicity.  We look
for solutions of the time-dependent Schr\"{o}dinger equation:
$$
\ri \frac{\partial}{\partial t} \Psi (t) = H \Psi (t) 
$$
with the quantization condition: $\left[ x, p \right] = \ri $.  

There are two ways to solve this problem.  The easier one is the {\it
algebraic} method in which one introduces "creation" and
"annihilation" operators:
$$
a^\dagger   =  \frac{1}{\sqrt{2\omega}} \left( - \ri p + \omega x \right) ,  
\qquad
a~   =   \frac{1}{\sqrt{2\omega}} \left( \ri p + \omega x \right) 
$$
which satisfy a usual commutation relation: $\left[ a, a^\dagger
\right] = 1$ and diagonalize the Hamiltonian: $H = \omega \left(
a^\dagger a + \frac{1}{2} \right)$.  The ground state of the
Hamiltonian is given by the condition: $a | 0 \rangle = 0 $.  This
method is usually transcribed to the quantization of fields and one
obtains particle creation and annihilation operators as basic building
blocks in describing physical processes.  In this approach, one
usually does not refer to the wave function of the system explicitly
but instead concentrates on amplitudes of particular process for given
initial and final states specified by the particle number and other
quantum numbers.

Alternatively, one can solve the problem by the {\it analytic} method
in which one expresses the Schr\"{o}dinger equation in coordinate
representation with the differential operator $ p = - \ri \partial /
\partial x$ and solve the resultant second order differential
equation.  The eigenfunction of this differential equation are
Hermite's polynomials with the Gaussian ground state wave function:
\begin{equation}
\Psi_0 (x) = \langle x | 0 \rangle = 
\left( \frac{\omega}{\pi} \right)^{1/4} e^{- \frac{1}{2}\omega x^2} 
\end{equation}
It is straightforward to write down the ground state wave functional of
the quantized free field $\varphi (x)$.  For one component scalar
field theory, it becomes just a product of the Gaussian wave functions
of normal modes each specified by the momentum with the oscillator
frequency $\omega_k = \sqrt{m^2 +k^2}$:
$$
\Psi_0 [ \varphi (x) ] = 
{\cal N} \exp \left[ - \frac{1}{2} \sum_k \omega_k \varphi_k^* \varphi_k \right].
$$
with a proper normalization condition.

\vskip 10pt
\leftline{\bf Time-dependent variational wave function:}
\vskip 5pt Now let us consider time-dependent solutions of the
harmonic oscillator.  We first modify the ground state wave function by
adding extra complex phase factor $e^{\ri \left( p_0 - i\omega x_0
\right) x }$.  One then obtains the Gaussian wave function with its the
center shifted
\begin{equation}\label{sgauss}
\Psi (x, t; x_0, p_0) %
= \exp \left[ - \frac{1}{2} \omega \left( x - x_0 (t) \right)^2 
+ \ri p_0 (t) x \right]
\end{equation}
Here $p_0 (t), x_0 (t)$ are time-dependent parameters which are to be
determined by imposing that the above function is a solution of the
time-dependent Schr\"{o}dinger equation.  Here we use the variational
method to derive the equations of motion of $p_0 (t), x_0 (t)$.

The Schr\"{o}dinger equation can be obtained by imposing a stationary
condition $\delta S = 0$ for the action:
$$
S [\Psi] = \int dt \langle \Psi (t) | H - \ri \frac{\partial}{\partial
t} | \Psi (t) \rangle
$$
with respect to variation of the wave function $\langle \Psi (t) |$.
For the variational wave function in the form of \ref{sgauss}, the
integrand can be computed easily with $\langle \Psi | p^2 | \Psi
\rangle = p_0^2$, $\langle \Psi | x^2 | \Psi \rangle = x_0^2$,
$\langle \Psi | i\frac{\partial}{\partial t} | \Psi \rangle = -
\dot{p}_0 x_0$:
\begin{equation}
\langle \Psi | H - i\frac{\partial}{\partial t} | \Psi \rangle 
=  p_0^2 + \omega^2 x_0^2 + \dot{p}_0 x_0 .
\end{equation}
Taking stationary conditions of the action with respect to two
time-dependent parameters, $\delta S / \delta p_0 = \delta S / \delta
x_0 = 0$, one finds
\begin{equation}
\dot{p}_0 = - \omega^2 x_0 , \qquad \dot{x}_0 = p_0 
\end{equation}
which are just the {\it classical} equation of motion of the harmonic
oscillator.  The center-shifted Gaussian wave function may be obtained,
beside the phase factor $e^{- \ri p_0 x}$, by operating a translation
operator on the ground state:
$$
| 0; x_0, p_0 \rangle \sim \exp [ -\ri x_0 p ] | 0 \rangle = 
\exp \left[ -\ri \sqrt{\frac{\omega}{2}} x_0 (a^\dagger - a ) \right] | 0 \rangle . 
$$  
It thus describes a phase-coherent mixture of infinite number of
excited states $| n \rangle$.  The corresponding states in quantum
field theory are called {\it coherent states}.  One thus sees that
classical field equations arises if one describes the wave functional
of the quantized fields in terms of coherent states.

Keeping Gaussian form of variational wave function one can go one step
further by introducing additional phase factor quadratic in $x$.  The
wave function now takes the form of
\begin{equation}\label{squeezed}
\Psi_{\rm sq.} (x,t) = \left( \frac{\mu}{\pi} \right)^{1/4}
\exp \left[ 
- \frac{1}{2} (\mu + i\sigma  ) ( x - x_0)^2 - \ri p_0 x  \right]
\end{equation}
where we have introduced two new real parameters, $\mu, \sigma$,
characterizing the width of the wave function, which we shall consider
as time-dependent variational parameters.  The time-dependent
parameter $\mu (t)$ is related to the quantum fluctuation of the
position of the particle around its mean $x_0 = \langle x \rangle$ by
$\langle (x - x_0)^2 \rangle = 1/\mu (t)$.  The another parameter
$\sigma$ is related to the rate of change of $\mu$ as we shall see
below.  This modification of the wave function thus describes the
breathing motion (squeezing and stretching) of the wave function
centered at $x_0 (t)$.  In the quantum field theory, these generalized
coherent states are called {\it squeezed states}.

With the new variational wave function the action integrand becomes
$$
\langle \Psi (t) | H -\ri \frac{\partial}{\partial t}| \Psi (t) \rangle_{\rm sq.} = 
\frac{1}{2} \left( p_0^2 + \omega^2 x_0^2 \right) 
+ \frac{1}{4} \left( \mu + \frac{\sigma^2}{\mu} +
 \frac{\omega^2}{\mu} \right) 
+ \dot{p}_0 x_0 - \frac{\dot{\sigma}}{4\mu}
$$
Taking the stationary conditions with respect to the variations of
$\mu (t), \sigma (t)$, in addition to classical parameters, $x_0 (t),
p_0 (t)$, one finds
\begin{equation}\label{mu-sigma}
\dot{\mu} = 2 \sigma \mu, \qquad \dot{\sigma} = \sigma^2 + \omega^2 - \mu^2, 
\end{equation}
while the equations of motion of $x_0 (t), p_0 (t)$ are unchanged.  We
see that the imaginary part $\sigma$ of the Gaussian width parameter
plays a role similar to the velocity of the motion of the center of
the Gaussian $x_0$.  For the harmonic oscillator Hamiltonian, coherent
states and squeezed states are exact solutions of the Schr\"{o}dinger
equations. The classical motion and the quantum fluctuation decouple
in this exactly soluble problem.  This is not the case, however, when
the potential is not quadratic in $x$.

\vskip 10pt
\leftline{\bf Anharmonicity:}
\vskip 5pt 

To illustrate the effect of non-harmonic part of the potential, we add
a term quartic in $x$ in our Hamiltonian:
$$
H_I = 
\frac{\lambda}{4!} x^4 
$$
In the scalar field theory, this corresponds to adding a $\varphi^4$
self-interaction term.  This anharmonic term generates new terms in
the integrand of the action:
$$
\langle \Psi (t) | H_I | \Psi (t)  \rangle_{\rm sq.}  = 
\frac{\lambda}{4!}  \left( \frac{3}{4\mu^2} + 3 \frac{x_0^2}{\mu} + x_0^4 \right)
$$
which cause coupling between the classical motion of the mean of
particle position and the quantum fluctuation around it:
\begin{eqnarray*}
\dot{x}_0  = p_0, &\qquad&
\dot{p}_0  = 
- \omega^2 x_0 - \frac{\lambda}{6}x_0^3 - 
\frac{\lambda}{4\mu}x_0, \\
\dot{\mu}  =  2 \sigma \mu, &\qquad&
\dot{\sigma} =  \sigma^2 + \omega^2 - \mu^2 
+ \frac{\lambda}{4\mu}+ \frac{\lambda}{2}x_0^2
\end{eqnarray*}

Let us see how the time independent solution (the ground state) is
modified by the anharmonic term.  The conditions $ \dot{x}_0 =
\dot{\mu}_0 = 0$ demand that $p_0 = \sigma =0$, while the remaining
two conditions $ \dot{p}_0 = \dot{\sigma}_0 = 0$ determine the values
of $x_0$ and $\mu$.  There are two types of solutions depending on the
sign of $\omega^2$.  When $\omega^2 > 0$ we have a "normal" solution
centered at the origin $x_0 = 0$ but the width is modified slightly as
determined by:
$$
\mu^2 = \omega^2 + \frac{\lambda}{4\mu}
$$
This equation may be called the {\it gap equation} for the reason to
be discussed in the next section.  If $\omega^2 < 0$, the potential
has double minima at $x_{\rm min.} = \pm \sqrt{ 6 / \lambda} \omega$
and the $x_0 = 0$ point becomes local maximum of the potential.  In
this case, there are two "symmetry breaking" solutions centered at the
two solutions of
$$
\omega^2  + \frac{\lambda}{4\mu} + \frac{\lambda}{6}x_0^2 = 0 
$$
where the value of the width parameter $\mu$ is determined
self-consistently with the modified gap equation:
$$
\mu^2 = - 2 \omega^2 - \frac{\lambda}{2\mu} .
$$
We note that the position of the center of our Gaussian variational
wave function shift slightly toward the origin from the position of the
minimum of the potential.

\section{Scalar Field Theory: $\varphi^4$ model}

\newcommand{\ba}{\begin{array}} 
\newcommand{\ea}{\end{array}}
\newcommand{\bc}{\begin{center}}
\newcommand{\ec}{\end{center}}
\newcommand{\hg}{{\hat G}}
\newcommand{\hga}{{\hat \Gamma}}
\newcommand{\hsig}{{\hat \Sigma}}
\newcommand{\baro}{{\bar \rho}}
\newcommand{\la}{\langle}
\newcommand{\ra}{\rangle}
\newcommand{\bpi}{\bar \pi}
\newcommand{\bphi}{\bar \phi}
\newcommand{\bvarphi}{\bar \varphi}

Having been warmed up by a much simpler problem in quantum mechanics,
it is now our task to transcribe the result to a problem in quantum
field theories.  We consider first a prototype scalar field theory
with $\varphi^4$ self-interaction.

The Hamiltonian density of the theory is given by
\begin{equation}\label{2e1}
{\cal H}({\bf x}) = \frac{1}{2} \pi^2 ({\bf x}) + \frac{1}{2} \left(
\nabla \varphi ({\bf x}) \right)^2 + \frac{m_0^2}{2} \varphi^2 ({\bf
x}) + \frac{\lambda}{24} \varphi^4 ({\bf x}) \ , 
\end{equation} 
where $\pi ({\bf x})$ is an operator conjugate to the field $\varphi
(x)$ and is expressed by a functional derivative $-\ri \delta/\delta
\varphi ({\bf x})$.
\footnote{A care must be taken to give precise mathematical meanings for these
expressions\cite{ML85}, but we do not go into such problems here. }
We write a Gaussian time-dependent variational wave functional 
for this Hamiltonian formally as
\begin{equation}\label{2e2} 
\Psi \left[ \varphi ({\bf x}) \right] = {\cal N} \exp \left( \ri \la
{\bar \pi} | \varphi - {\bar \varphi} \ra - \la \varphi- \bar \varphi
| \frac{1}{4G} + \ri \Sigma | \varphi- \bar \varphi \ra \right) ,
\end{equation}
where $G$, $\Sigma$, $\bar \varphi$, ${\bar \pi}$ define respectively
the real and imaginary part of the kernel of the Gaussian width and
its average position and momentum.  We have used the short hand
notation 
$ \la {\bar \pi} |{\varphi} \ra = \int {\bar \pi}({\bf x}, t )
\varphi({\bf x}) d{\bf x}.$
Although it looks a little horrible, it is just a straightforward
generalization of \eqref{squeezed}.  The correspondences between the
previous quantum mechanical example and the present case are
summarized in the Table 1.

\begin{table}[t]
\begin{center}
\caption{Correspondence between quantum mechanics and scalar field theory}
\vskip 10pt
\begin{tabular}{| c | c |}
\hline
\hline
quantum mechanics & scalar field theory \\
\hline
$ x$ & $ \varphi ({\bf x})$ \\
$ \omega^2$ & $m^2_0 - {\bf \Delta} $ ( or $m^2_0 + {\bf k}^2$ ) \\
$ \frac{\lambda}{4} x^4$ & $ \frac{\lambda}{4}\varphi^4 ( {\bf x} )$ \\
$x_0 (t), p_0 (t)$ & $ {\bar \varphi} ( {\bf x}, t ),  {\bar \pi}( {\bf x}, t )$ \\
$\mu (t) , \sigma (t)$ & $\frac{1}{2}G^{-1} ( {\bf x}, {\bf y}, t), 2 \Sigma ( {\bf x}, {\bf y}, t) $ \\
$ \mu \xi$ & $G^{-1/2} \xi G^{-1/2}$ \\ 
\hline
\hline
\end{tabular}
\end{center}
\end{table}


The equations of motion are found to be
\begin{equation}\label{2e4}
\begin{array}{lll}
\displaystyle
\dot {\bar \varphi } & = & -\bar \pi ,
\\
& \\
\dot {\bar \pi} & = & 
\left( -\Delta + m_0^2 + \frac{\lambda}{6} \bar \varphi^2 ( {\bf x} ) +
\frac{\lambda}{2}  G({\bf x}, {\bf x}) \right) \bar \varphi, \\
& \\
\dot G  & = & 2 (G \Sigma +\Sigma G), \\
& \\
\dot \Sigma & = & \frac{1}{8} G^{-2} - 2\Sigma^2 -
\frac{1}{2} \left( -\Delta + m_0^2 + \frac{\lambda}{2} \bar \varphi^2 +
\frac{\lambda}{2}  G({\bf x},{\bf x}) \right) .
\end{array}
\end{equation}
where it is understood that ${\bar \varphi }$ and ${\bar \pi}$ denote
vectors with components ${\bar \varphi } ({\bf x}) $ and ${\bar \pi}
({\bf x})$, $G $ and $\Sigma$ denote matrices with matrix elements $G
({\bf x}, {\bf y})$ and $\Sigma ({\bf x}, {\bf y})$, and the matrix
product $G \Sigma $ is given by $ \int d{\bf z} G ( {\bf x}, {\bf z} )
\Sigma ( {\bf z}, {\bf y} ) $.

For the vacuum, we should have time-independent solution so that
${\bar \pi} = \Sigma = 0$, ${\bar \varphi} ({\bf x} ) = \varphi_0$,
and
\begin{equation}\label{2e6}
G ( {\bf x} , {\bf y}, t ) = G_0 ( {\bf x} -  {\bf y} ) = 
\int d {\bf k}  \frac{e^{ \ri ( {\bf x} -  {\bf y} ) \cdot {\bf k} }}{2 \sqrt{\mu^2 + k^2}}.
\end{equation}
where the effective mass $\mu$ is determined self-consistently by the
 non-linear integral equation
\begin{equation}\label{2e7}
\mu^2= m_0^2 + \frac{\lambda}{2} G_0 ( 0 )  + \frac{\lambda}{2} \varphi_0^2 \ .
\end{equation}
which is usually called the gap equation because it determines the
mass gap self-consistently.  In the symmetric phase the expectation
value of the field $\varphi_0$ vanishes while in the symmetry broken
phase it must be such that
\begin{equation}\label{2e8}
m_0^2 + \frac{\lambda}{2} G_0 ( 0 )  + \frac{\lambda}{6} \varphi_0^2=0.
\end{equation}
This last equation implies that $\mu^2$=$\lambda \varphi_0^2/3$.  The
equations of motion \eqref{2e4} determine the time-evolution of the
variational wave functional \eqref{2e2}.

We comment on two well-known problems which are absent in quantum
mechanics but are characteristic in quantum field theory: divergences
and covariance.  In the above expression the momentum integral in $G_0
(0)$ is quadratically divergent. 
As well-known, this divergence originates from
couplings of infinite degrees of freedom (all momentum modes) in the
interaction of local fields.  In
the renormalizable field theories, these divergences (or cut-off
dependence) may be absorbed into an appropriate redefinition of finite
number of physical parameters, such as the mass and the coupling
constant.  This renormalization procedure works at least order by
order in the power series expansion in terms of the coupling constant.
In our non-perturbative calculation scheme with the variational
method, we may absorb the cut-off dependence into the mass and the
coupling constant.\cite{KV89,PS87} We encounter, however, well-known
triviality problem: renormalized coupling constant $\lambda_R$ becomes
zero as one sends the cut-off $\Lambda$ to infinity keeping the
original coupling $\lambda$ positive for the stability of the ground
state.  Despite this well-known pathology, the model still makes sense
as an effective theory with the finite cut-off.

The another problem is the lack of manifest covariance in our
formulation: the time coordinate $t$ have been treated differently
from spatial coordinates throughout calculations.  This is an old
problem which was originally solved in QED by Tomonaga, Schwinger,
Feynman and Dyson who developed the covariant perturbation theory
based on the interaction representation.  Vautherin invented a new
ingenious trick to derive manifest covariant form of the equations of
motion which I shall now describe. \cite{VM97,TVM99}

\vskip 10pt
\leftline{\bf Mean field equations in the Hartree-Bogoliubov form:}
\vskip 5pt
A key ingredient of his method is the reduced density matrix defined by
\begin{equation}\label{3a1}
{\cal M}({\bf x},{\bf y};t) =
\left( \ba{cc} 
\ri \la {\hat \varphi}({\bf x}) {\hat \pi}({\bf y}) \ra -1/2
&  \la {\hat \varphi}({\bf x}) {\hat \varphi}({\bf y}) \ra  \\ 
\la {\hat \pi}({\bf x}) {\hat \pi}({\bf y}) \ra
&-\ri \la {\hat \pi}({\bf x}) {\hat \varphi}({\bf y}) \ra  -1/2
\ea \right),
\end{equation}
where ${\hat \varphi}=\varphi-{\bar \varphi}$ , ${\hat \pi}=\pi-{\bar \pi}$,
and expectation values are calculated with the Gaussian functional
$\Psi(t)$ and is given by
\begin{equation}\label{3a2}
{\cal M}=
\left( \ba{cc}  -2iG \Sigma  &  G \\ 
\frac{1}{4} G ^{-1} + 4 \Sigma G \Sigma  & 2\ri \Sigma G  \ea \right).
\end{equation}
Using \eqref{2e4}, one can show that the equation of motion of the
reduced density matrix can be cast into the Liouville-von Neumann form
\begin{equation}\label{LvN}
\ri {\dot {\cal M}}= [{\cal H}, {\cal M}],
\end{equation}
where the generalized Hamiltonian ${\cal H}$ is given by
\begin{equation}\label{3a13}
{\cal H}= \left( \ba{ll} 0 &1 \\ \Gamma & 0  \ea \right).
\end{equation}
with 
\begin{equation}\label{3a11}
\Gamma= -\Delta + m_0^2 + \frac{\lambda}{2} \bar \varphi^2 +
\frac{\lambda}{2}  G({\bf x},{\bf x}).
\end{equation}
This form of equations is known in many-body theory as the time-dependent 
Hartree-Bogoliubov equations.   

The reduced density matrix satisfies ${\cal M}^2= \frac{1}{4}{\bf I}$
so that it has two eigenvalues of $\pm 1/2$.  We write the $n$-th
eigenvector of ${\cal M}$ with eigenvalue $1/2$ as $(u_n ({\bf x}) ,
v_n ({\bf x}) )$:
\begin{equation}\label{3a4}
{\cal M}
\left( \ba{l} u_n \\ v_n \ea \right)=  \frac{1}{2}
\left( \ba{l} u_n \\ v_n \ea \right) ,
\end{equation}
then one can show that vectors $ (u^*_n ({\bf x}), -v^*_n ({\bf x})) $
give eigenvectors for eigenvalue $-1/2$.  The $u$ and $v$ components
of eigenvectors are called {\it mode functions}.  With these
eigenvectors the reduced density matrix has a spectral decomposition
as
\begin{equation} \label{3a7}
{\cal M}=
\frac{1}{2} \sum_{n>0} \left[ 
\left( \ba{l} u_n \\ v_n \ea \right) \left( v_n^*, ~u_n^* \right)+
\left( \ba{l} ~~ u_n^* \\ -v_n^* \ea \right) \left( -v_n, ~u_n \right)
\right].
\end{equation}

The Liouville-von Neumann equation \eqref{LvN} can be rewritten in terms 
of the mode functions as 
\begin{equation}\label{4a1}
\ri \partial_t \left( \ba{l} u_n \\ v_n \ea \right)=
\left( \ba{ll} 0 &1 \\ \Gamma &0  \ea \right)
\left( \ba{l} u_n \\ v_n \ea \right).
\end{equation}
Eliminating $v_n$ and inserting \eqref{3a11} we obtain a modified
Klein-Gordon type equation for the mode functions
\begin{equation}\label{4a2}
\left( \Box + m_0^2 + \frac{\lambda}{2} \bar \varphi^2 +
\frac{\lambda}{2}  G({\bf x},{\bf x}) \right) u_n =0 ,
\end{equation}
where the spectral representation \eqref{3a7} implies
$$
G({\bf x},{\bf x}) = \la {\bf x}| G(t) |{\bf x} \ra = \frac{1}{2} \sum_n | u_n({\bf x},t)|^2.
$$
To write the equation of motion fully covariant way, we introduce a
Feynman propagator in terms of the mode functions
\begin{equation}\label{4a12} \ba{lll}
\la x | S | y \ra & = &
\theta(x_0 - y_0 ) \sum_{n>0} u_n^*({\bf x},x_0)  u_n({\bf y},y_0)  \\
&~~\\ 
& & \qquad 
+ ~~ \theta(y_0 - x_0 ) \sum_{n<0} u_n^*({\bf x},x_0)  u_n({\bf y},y_0).
\ea \end{equation}
so that 
\begin{equation}\label{4a13}
\la {\bf x}| G(x_0)| {\bf x} \ra 
= \la x | S | x \ra 
\end{equation}
Then we finally arrive at a fully covariant, self-consistent equations
of motion for the mode functions\cite{VM97,TVM99} which reproduces the
same equations obtained earlier by the functional integral
method.\cite{CJT74}

This way of writing the equations of motion also paves a way to generalize 
the calculation at finite temperatures. 

\section{Statistical Ensembles}
Foregoing discussions are limited to evolution of a single coherent
(Gaussian) state.  In realistic physical situations, we are more
interested in the evolution of the statistical ensemble which is
described by the density matrix:
\begin{equation}
{\hrho} (t) = \sum_n | \Psi_n (t) \rangle p_n (t) \langle \Psi_n (t) |
\end{equation}
where $p_n (t)$ is a probability distribution specifying the ensemble
hence It should satisfy $\sum_n p_n (t) = 1$.  The expectation value
of an observable $O$ is given by $\langle O \rangle = {\rm Tr} {\hrho}
O $ and the statistical entropy $S $ is given by
\begin{equation}
S  = - {\rm Tr} {\hrho} \ln {\hrho} = - \sum_n p_n \ln p_n
\end{equation}
Hence if $p_n$ is time-independent then the entropy is conserved.  In
equilibrium at temperature $T = 1/\beta$, the density matrix is given
by the canonical ensemble: $p_n^{eq.} (t) = e^{-\beta E_n } / Z$ where
$Z = \sum_n e^{- \beta E_n} $ which maximizes the entropy $S(t)$ under
the condition of fixed expectation value of the energy $E = \langle H
\rangle$.  The density matrix with time-independent $p_n$ obeys the
Liouville-von Neumann equation: $ \ri \partial_t \hrho = [ H, \hrho ]
$.

The variational method has been extended for the time-dependent
density matrix by Eboli, Jackiw and Pi. \cite{EJP88} Without going
into detail, we illustrate the essence of their method in terms of
simple quantum mechanical example with harmonic oscillator Hamiltonian
\eqref{ho}.  We first introduce the coordinate representation of the
density matrix by
\begin{equation}
\rho ( x, y ; t ) = \langle x | {\hrho} (t) | y \rangle =
\sum_n  \Psi_n ( x, t)  p_n (t) \Psi_n^* ( y, t) 
\end{equation}
and observe that with single Gaussian variational wave function the
density matrix is just a product of two Gaussian: $\rho ( x, y ; t )
\sim e^{ - \omega ( x^2 + y^2 ) / 2} $.  One can show that in the
other extreme limit of thermal equilibrium, the density matrix $\rho (
x, y ; t ) $ can also be expressed by a mixed Gaussian form.  This is
so because the equilibrium density matrix ${\hat \rho}_{eq.} (\beta) =
e^{- \beta H}$ obeys the imaginary time Schr\"{o}dinger equation (the
Bloch equation): $ - \partial_\tau {\hat \rho}_{eq.} ( \tau ) = H
{\hat \rho}_{eq.} $.  Using $\rho _{eq.} ( x, y ) = \langle x | {\hat
U} ( -\ri \tau ) | y \rangle$ and well-known path-integral expression
of the matrix elements of the unitary evolution operator ${\hat U} (t)
= e^{-\ri t H}$, one finds
\begin{equation}\label{rhoeq}
\rho _{eq.} ( x, y ) =
\left( \frac{\omega}{2 \pi \sinh \omega \beta} \right)^{1/2}
\exp \left[ - \frac{\omega}{2 \sinh \omega \beta} 
\left\{ (  x^2 + y^2 ) \cosh \omega \beta - 2 x y\right\} \right]
\end{equation}
which is again Gaussian with an extra term containing a product of two
coordinates $xy$.

For more general mixed states we introduce a generalized Gaussian
density matrix\cite{EJP88}
\begin{equation}\label{grho}
\rho ( x, y ; t) = {\cal N} \exp \left[ \ri p_0 ( \hx - \hy ) - 
\frac{\mu}{2} \left( \hx^2 + 
\hy^2  - 2 \xi  \hx \hy \right) + \ri \frac{\sigma}{2} ( \hx^2 - \hy^2 ) \right]
\end{equation}
where we have used the time-dependent shifted coordinates: $\hx = x -
x_0 $ and $\hy = y - x_0$.  Equations of motion of the time-dependent
parameters of the generalized Gaussian density matrix \eqref{grho} can
be derived from the Liouville equation of the density matrix and we
obtain a set of equation similar to \eqref{mu-sigma} with a small
modification:
\begin{equation}\label{mu-sigma}
\dot{\mu} = 2 \sigma \mu, \quad 
\dot{\sigma} = \sigma^2 + \omega^2 - ( 1 - \xi^2) \mu^2.
\end{equation}
The parameter $\xi$ is called the mixing parameter which measures the
degree of mixture of different pure states in the ensemble; it remains
constant for an adiabatic evolution of the system.  In equilibrium,
$p_0 = x_0 = \sigma = 0$ and other two parameters are given by the
specific functions of temperature: \begin{equation}
\mu_{eq.} = \omega \coth \omega \beta ,
\quad
\xi_{eq.} = \cosh^{-1} \omega \beta
\end{equation}
as indicated by the formula \eqref{rhoeq}.  Extension of the Gaussian
density matrix in quantum mechanics to that in quantum field theories
is straightforward as indicated in the last row of the Table 1.

The reduced density matrices we have introduced in the previous
section for a pure Gaussian state can be extended for a mixed state
immediately by the replacements:
\begin{equation}
\la {\hat \varphi}({\bf x}) {\hat \pi}({\bf y}) \ra 
= {\rm Tr} ( \hrho (t) {\hat \varphi}({\bf x}) {\hat \pi}({\bf y}) ), 
\la {\hat \varphi}({\bf x}) {\hat \varphi}({\bf y}) \ra 
= {\rm Tr} ( \hrho (t) {\hat \varphi}({\bf x}) {\hat \varphi}({\bf y}) ), {\rm etc.}
\end{equation}
and one can derive equations of motion of the reduced density matrix
similar to the pure state case.  In the case of equilibrium
distribution, this amounts to introduce a factor containing the
occupation number in the sum over the mode functions.
 
\section{Rotating chiral condensate in the $O(N)$ sigma model}
We briefly mention about an application of the above method to the
$O(N)$ sigma model which is composed of $N$-components coupled scalar
fields $\varphi_n$ with continuous $O(N)$ symmetry.  We expect that
this global symmetry of the model is broken spontaneously at low
temperatures, characterized by non vanishing expectation value of one
component of fields, say ${\bar \varphi}_0 = {\rm Tr} {\hat \rho}
{\hat \varphi} \neq 0 $, and the system exhibits an order-disorder
phase transition to a state with ${\bar \varphi}_0 = 0$.  This is what
is expected in QCD where the chiral symmetry is broken in vacuum and
is expected to be restored at finite temperature.  This chiral phase
transition has been studied by an effective theory with pion and sigma
fields with $O(4)$ global symmetry.

We have applied our method to describe a special kind of
time-dependent condensate which rotates in a subspace of the internal
symmetry space:
\begin{equation}
\left( \ba{c} {\bar \varphi}_1 (x) \\ {\bar \varphi}_2 (x) \ea \right) = 
\left( \ba{cc} \cos (q \cdot x) & \sin (q \cdot x) \\ 
- \sin (q \cdot x) & \cos (q \cdot x) \\
 \ea \right) \left( \ba{c} \varphi_0 \\ 0 \ea \right) 
 = \exp [ \ri (q \cdot x) \tau_2 ]  \left( \ba{c} \varphi_0 \\ 0 \ea \right) 
\end{equation}
where $\tau_2 = \left( \ba{cc} 0 & -\ri \\ \ri & 0 \ea \right)$ and
$q_\mu = (\omega, {\bf q})$ is a four vector specifying the direction
of rotation: for a spatially uniform rotation in time it is a pure
time-like vector; while a static condensate with oscillation in space,
it is a pure space-like vector.  The matrix $U(x) = \exp [ \ri (q
\cdot x) \tau_2 ] $ can be considered as a "gauge transformation" to
the local rest frame of the rotating condensate.  In the rotating
frame, the four derivative which appear in the equation of motion of
the mode functions is "gauge transformed" to $U \partial_{\mu}
U^\dagger = \partial_{\mu} - \ri q_{\mu} \tau_2$ and the effect of the
rotation is seen in this frame as an appearance of the apparent
"centrifugal force".  Indeed, our result of phase diagram for such
dynamical condensates shows that the amplitude of the chiral
condensate with uniform time-like rotation increases due to the
centrifugal force.  This effect has been observed in the classical
solutions of Anselm and Ryskin \cite{AR91}; our quantum generalization
of their solution shows that this effect is amplified by the coupling
of quantum fluctuations to rotations, which the static condensate with
spatial oscillations are more suppressed by the quantum
fluctuations.\cite{VM98} A phase diagram of rotating condensate was
obtained in \cite{TVM99} and the damping of the rotating condensate
due to the symmetry breaking perturbation was computed by the method
of the response function in \cite{TVM00}.

\section{Outlook}
Vautherin started to work on the variational approach to quantum field
theories many years ago with Arthur Kerman.  They developed many
important ideas in their unpublished works and tried to solve QCD
non-perturbatively with their method with a hope to gain new insights
in the quark confinement problem.\cite{KV89} The Gaussian Ansatz for
the variational wave functional however has difficulty of breaking the
local gauge invariance and the projection to color singlet state
destroys a nice feature of the Gaussian wave functional.\cite{KK95}
Vautherin continued to work on the problem with his students and
brought a new insight into the problem again introducing a technique
developed in nuclear many-body theory in his last paper.\cite{HIMV00}
His efforts in this direction may be carried over to study the
dynamical evolution of the quark-gluon plasma in ultrarelativistic
nucleus-nucleus collisions.  Vautherin was also interested in the
recent experimental breakthrough of creating weakly interacting
Bose-Einstein condensates in well-controlled laboratory environments.
Our method can be also applied to such problem to investigate the
effect of quantum fluctuations which are usually ignored in
theoretical descriptions.\cite{TVM01}

\vskip 10pt 

I am much indebted to Yasuhiko Tsue as well as to Dominique Vautherin
for our works quoted above.  I thank them for sharing the joy of
physics in our collaboration.


\end{document}